\documentclass[preprint,pra,superscriptaddress,showpacs,amsmath]{revtex4}

\usepackage{graphicx}

\begin{document}

\title
{Quantum Optics with Quantum Gases}
\date{Submitted October 7, 2008}

\author{I. B. Mekhov}
\email{Mekhov@yahoo.com}
\affiliation{Institut f\"ur Theoretische Physik, Universit\"at
Innsbruck, Innsbruck, Austria} \affiliation{St. Petersburg State
University, Faculty of Physics, St. Petersburg, Russia}

\author{H. Ritsch}
\affiliation{Institut f\"ur Theoretische Physik, Universit\"at
Innsbruck, Innsbruck, Austria}

\begin{abstract}
Quantum optics with quantum gases represents a new field, where the
quantum nature of both light and ultracold matter plays equally
important role. Only very recently this ultimate quantum limit of
light-matter interaction became feasible experimentally. In
traditional quantum optics, the cold atoms are considered
classically, whereas, in quantum atom optics, the light is used as
an essentially classical axillary tool. On the one hand, the
quantization of optical trapping potentials can significantly modify
many-body dynamics of atoms, which is well-known only for classical
potentials. On the other hand, atomic fluctuations can modify the
properties of the scattered light.
\end{abstract}


\maketitle

\newpage

\section{Introduction}

The general goal of this direction of our research is to join two
broad and intensively developing fields of modern physics: quantum
optics and ultracold quantum gases. Such a unified approach will
consider both light and matter at an ultimate quantum level, the
level that only very recently became accessible experimentally
\cite{Exp}.

Optics has become one of the most well-established and exciting
disciplines in physics. Even classical optics, treating the light as
classical electromagnetic waves, led to the important discoveries
and technological breakthroughs. For example, up to now, the optical
measurements provide us the highest level of precision. A new era in
optics started in the 20th century with the creation of quantum
theory and invention of a laser, when the concept of photons came
into existence and became testable experimentally. Now, quantum
optics, which studies nonclassical light (i.e., the light whose
properties cannot be explained by classical optics, but well
described using the quantum wave-particle dualism), also achieved a
very high level of development.

In the last decades of the 20th century, the progress in laser
cooling of atoms led to the foundation of a new field in atomic
physics: atom optics. According to quantum mechanics, at low
temperatures, massive particles behave similarly to the waves with
the wavelengths larger for lower temperatures. Thus, the matter
waves in atom optics behave analogously to the light waves in
optics. The quantum properties of matter waves became well
accessible, after the ultralow temperatures were reached and the
first Bose-Einstein condensation (BEC) was achieved in 1995. An
exiting demonstration of "quantum atom optics" was presented in
2002, when the phase transition between two different quantum states
of matter waves (superfluid and Mott insulator) was achieved.

Thus, the role of light and matter in quantum atom optics and
quantum optics is totally reversed. However at present, even in the
most involved works, the role of light in quantum atom optics is
essentially reduced to a classically auxiliary tool.  One can create
and manipulate intriguing atomic quantum states using the forces and
potentials produced by classical light waves. For example, the light
beams are used to form beam splitters, mirrors, and other "devices"
known in optics, but now applied for matter waves. In this context,
the periodic micropotentials (the famous optical lattices) are
analogies of cavities in classical optics, as they enable storing
and manipulating atomic states.

Quantum optics with quantum gases will close the gap between
standard quantum optics and quantum atom optics. In contrast to the
traditional works, it will address the phenomena, where the quantum
nature of light and ultracold gases is equally important. So, the
quantum optics with quantum gases can be considered as an ultimate
quantum limit of light-matter interaction, which became
experimentally feasible only recently, when a BEC was coupled to the
light mode of a cavity \cite{Exp}.

As both light and atoms will be considered at the quantum level,
quantum optics with quantum gases will enable the unprecedented
control of light and matter. This will find applications in the
following areas. Novel non-destructive detectors of atomic states
using light scattering (currently unavailable). Quantum information
processing: novel protocols will be developed using the multipartite
entangled states naturally appearing at this level of interaction.
Quantum interferometry and metrology: the multipartite entangled
states of massive particles are considered as a resource to approach
the ultimate Heisenberg limit, which can be used in the
gravitational wave detectors and novel quantum nanolithography.
Moreover, as the far off-resonant interactions can be considered,
the general approaches can be applied to other fields: molecular
physics (quantum molecular gases were recently obtained);
semiconductor systems (BEC of exciton-polaritons); superconductor
systems (circuit cavity quantum electrodynamics).

A natural way to couple the quantum light and quantum gas is to load
ultracold atoms in a high-Q cavity. Theoretically, we contributed to
this filed \cite{NatPh,PRL,PRA,EPJD,arxiv,Andras,PRL05}, which has
stimulated further theoretical research in other groups as well
\cite{otherTh}.

On the one hand, the properties of scattered light will be affected
by the quantum characteristics of the atomic states, e.g., by the
particular atom number fluctuations. Thus the light can serve as a
non-destructive probe of the atomic state. The aim of this paper is
to summarize our results on light scattering and outline some
perspectives by presenting several particular cases and using simple
physical models and interpretations.

On the other hand, the quantum nature of trapping potentials
provided by the cavity field can significantly modify the many-body
dynamics and quantum phase transitions, well known only for
classical potentials. Moreover, the interaction between particles
via a cavity field provides a new type of the long range
correlations that has not been studied in the standard condensed
matter problems, and can also lead to novel strongly correlated
systems. This part of the problem is outside of the scope of the
present paper (for the generalized Bose-Hubbard model taking into
account the quantized potentials see \cite{PRA,EPJD}).

\section{Theoretical model}

We consider $N$ two-level atoms trapped in a deep optical lattice
with $M$ sites formed by strong laser beams. A region of $K\le M$
sites is illuminated by two additional light modes (Fig.~1). We will
be interested in a situation, where one mode plays a role of the
probe, while another one represents the scattered light, which is
collected by a cavity and measured.

Interestingly, the many-body quantum problem can be significantly
simplified, if one assumes the atomic tunneling much slower than
fast light dynamics, which is a reasonable approximation. Then, the
full problem to describe light scattering reduces to simple
equations, which have a very transparent physical interpretation.
The interpretation has a direct classical analogy, keeping however
essentially quantum features.

The Heisenberg equation for the annihilation operator of the cavity
light mode $a_1$, where $a_0$ is the classical probe amplitude, with
the frequencies $\omega_{1,0}$ and mode functions $u_{1,0}({\bf r})$
is

\begin{eqnarray}\label{1}
\dot{a}_1= -i\left(\omega_1
+\frac{g^2\hat{D}_{11}}{\Delta_{1a}}\right)a_1
-i\frac{g^2\hat{D}_{10}}{\Delta_{1a}}a_0 -\kappa a_1+\eta(t), \\
\text{with} \quad \hat{D}_{lm}\equiv \sum_{i=1}^K{u_l^*({\bf
r}_i)u_m({\bf r}_i)\hat{n}_i},\nonumber
\end{eqnarray}
where $l,m=0,1$, $g$ is the atom-light coupling constant,
$\Delta_{la} = \omega_l -\omega_a$ are the large cavity-atom
detunings, $\kappa$ is the cavity relaxation rate, $\eta(t)=\eta
e^{-i\omega_{p}t}$ gives the external probe and $\hat{n}_i$ are the
atom number operators at a site with coordinate ${\bf r}_i$. We also
introduce the operator of the atom number at illuminated sites
$\hat{N}_K=\sum_{i=1}^K{\hat{n}_i}$.

In a classical limit, Eq.~(\ref{1}) corresponds to the Maxwell's
equation with the dispersion frequency-shifts of cavity mode
$g^2\hat{D}_{11}/\Delta_{1a}$ and the coupling coefficient between
the light modes $g^2\hat{D}_{10}/\Delta_{1a}$. For a quantum gas
those quantities are operators, which will lead to striking results:
atom number fluctuations will be directly reflected in such
measurable frequency- and angle-dependent observables. Thus, the
cavity transmission-spectra and angular distributions of scattered
light will reflect atomic statistics.

\section{Frequency dependences}

As a first example, we consider the simplest case: the atoms are
coupled to the single cavity mode (Fig.~2). Thus, we are interested
in the transmission spectra of a cavity around ultracold atoms. The
general Eq.~(\ref{1}) is reduced to

\begin{eqnarray}\label{2}
\dot{a}_1= -i\left(\omega_0
+\frac{g^2\hat{D}_{11}}{\Delta_{1a}}\right)a_0 -\kappa a_0+\eta(t),
\end{eqnarray}
which in the steady state has the solution for the cavity photon
number as

\begin{eqnarray}\label{3}
a^\dag_1a_1=\frac{|\eta|^2}{(\Delta_p-g^2\hat{D}_{11}/\Delta_{1a})^2
+\kappa^2},
\end{eqnarray}
where $\Delta_p=\omega_{p}-\omega_1$ is the probe-cavity detuning.
Theoretically, the simplest situation is realized for the
traveling-wave cavity (Fig.~2). In this case,
$\hat{D}_{11}=\hat{N}_K$. Thus the operator-valued frequency
dispersion shift, independently of the cavity-lattice angle, is
simply given by the atom number at $K$ illuminated sites, which is a
fluctuating quantity.

The transmission spectra are shown in Fig.~2. In the Mott insulator
state (MI), where all atom number fluctuations are totally
suppressed, one sees the classical transmission contour given by a
Lorentzian. In contrast, in the superfluid state (SF), where the
atom number fluctuations are strong, and $N_K$ can take any value,
one sees any possible Lorentzians with any possible dispersion
frequency shifts. The frequency distance between different
Lorentzians is given by a fluctuation produced by a single atom and
is equal to $g^2/\Delta_{1a}$.

One can show, that the transmission spectrum represents exactly the
atom number distribution function. Thus, the quantum phase
transition from the SF to MI is displayed in the transmission
spectrum as a shrinking to a single Lorentzian. Note, that for a
standing-wave cavity similar conditions can be easily analyzed
\cite{NatPh}.

Moreover, taking into account the quantum and dynamical nature of
the probe field, other atom number related quantities and their
distribution functions, e.g. the atom number difference between odd
and even sites \cite{NatPh}, can be accessed by the transmission
spectra. Interestingly, using that atom number difference one can
distinguish between the SF state (where the total atom number is
fixed) and the coherent-state approximation to the SF (where the
total atom number is infinite and thus unfixed). In SF, the
frequency distance between the Lorentzians will be twice as that in
the coherent state, since for the fixed atom number, the atom number
difference between odd and even sites changes with the step 2, while
for the coherent state it changes with the step 1. Thus for SF, the
splitting is $2g^2/\Delta_{1a}$, while for the coherent state it is
$g^2/\Delta_{1a}$.

\section{Angular distributions}

Let us now consider the configuration with the probe wave $a_0$.
However, in this section, we neglect the dispersion frequency shift
in Eq.~(\ref{1}), assuming it is smaller than the cavity relaxation
rate or probe-cavity detuning $\Delta_{01}$ (the consideration of
the external probe $\eta$ is also not necessary here). In this case,
the stationary solution is even simpler:

\begin{eqnarray}\label{4}
a_1=C\hat{D}_{10},
\end{eqnarray}
with the constant $C\equiv ig_0^2 a_0 /[\Delta_{1a}
(i\Delta_{01}-\kappa)]$. Thus, the light amplitude is proportional
to the coupling coefficient between two modes, which depends on the
atom numbers and, hence, is a fluctuating quantity.

Equation ~(\ref{4}) shows that the expectation value of the field
amplitude is simply proportional to $\langle\hat{D}_{10}\rangle$,
which depends on the mean atom numbers $\langle\hat{n}_i\rangle$.
Thus, if in different atomic states the mean atom number is the same
(as it is in MI and SF), the amplitude of the scattered light will
be also identical. However, the mean photon number $\langle
a^\dag_1a_1 \rangle$ is proportional to the second moment
$\langle\hat{D}_{10}^* \hat{D}_{10}\rangle$, which depends on the
density-density correlations $\langle\hat{n}_i\hat{n}_j\rangle$.
Those correlations are different for various atomic states, even if
the mean atom numbers are the same. Thus, in contrast to the light
amplitude, the photon number carries information about number
statistics in the atomic state.

We will demonstrate the difference in the light scattering from
different atomic states by considering the angle-dependent quantity
$R(\theta_0, \theta_1)$ (where $\theta_0$ and $\theta_1$ are the
angles between the lattice and light beams, cf. Fig.~1), which is
proportional to the difference ("noise") between the photon number
and classical intensity (the latter is just the amplitude squared):

\begin{eqnarray}\label{5}
R(\theta_0, \theta_1)\equiv \langle\hat{D}^*\hat{D} \rangle -
|\langle \hat{D} \rangle|^2 =\nonumber\\
 =\langle
\delta\hat{n}_a\delta\hat{n}_b\rangle \left|\sum_{i=1}^Ku_1^*({\bf
r}_i)u_0({\bf r}_i)\right|^2+(\langle\delta\hat{n}^2\rangle -
\langle
\delta\hat{n}_a\delta\hat{n}_b\rangle)\sum_{i=1}^K{|u_1^*({\bf
r}_i)u_0({\bf r}_i)|^2}.
\end{eqnarray}
Here we assumed that all pair correlations and on-site fluctuations
are the same for all lattice sites. For a 1D lattice of period $d$
and atoms trapped at $x_m=md$ ($m=1,2,...,M$) the mode functions are
$u_{0,1}({\bf r}_m)=\exp (imk_{0,1x}d)$ for traveling and
$u_{0,1}({\bf r}_m)=\cos (mk_{0,1x}d)$ for standing waves with
$k_{0,1x}=|{\bf k}_{0,1}|\sin\theta_{0,1}$ (cf. Fig.~1).

Thus, Eq.~(\ref{5}) shows that the difference between quantum and
classical light scattering $R(\theta_0, \theta_1)$ depends on the
on-site $\langle\delta\hat{n}^2\rangle$ and pair fluctuations
$\langle \delta\hat{n}_a\delta\hat{n}_b\rangle$. In MI, both type of
fluctuations are zero, so MI shows precisely classical light
scattering. In the SF state,
$\langle\delta\hat{n}^2\rangle=n(1-1/M)$ and $\langle
\delta\hat{n}_a\delta\hat{n}_b\rangle=-N/M^2$. Thus both
angle-dependent terms contribute to the difference. Note, that if SF
state is approximated by a coherent state, where the pair
fluctuations are neglected and $\langle\delta\hat{n}^2\rangle=n$,
only the second term contributes to the difference.

Figure 3 shows angular distributions for the intensity of classical
scattering (the lattice period is $d=\lambda_{0,1}/2$) and the
classical-quantum difference $R$ for the simplest configuration,
where both modes are traveling waves. In SF, except for the
isotropic background in $R$, the noise is suppressed at the
directions of the classical diffraction maxima. Such suppression
corresponds to the suppressed total atom number fluctuations in SF
state, in contrast to the coherent state (a totally isotropic
background is observed for the coherent state approximation).

Figure 4 shows similar situation, but the probe is the traveling
wave, whereas the cavity mode is the standing wave. Figure 5 shows
the situation, where both the probe and cavity are the standing
waves. One sees, that the noise displays the angular distribution
much richer than the classical diffraction. This is a consequence of
the second-order interference (the interference of intensities
similar to the Hanbury Brown and Twiss effect). The new features can
appear at the angles of the first order diffraction maxima, although
classically, only the zero order diffraction is possible.

Thus, the measurement of angular distribution of scattered light can
distinguish between different atomic quantum states (MI, SF, and
coherent in the above examples).

\section{Conclusions}

We have shown that various atomic quantum sates can be distinguished
by analyzing the frequency and angular distributions of the
scattered light, even if the mean atomic density is identical for
the atomic states. We have demonstrated the simplified model. This
model could be improved in the following directions. Here, the
trapping potentials are assumed to be very deep, so the atoms are
narrowly localized at each lattice site. However, one can take into
account the finite width of the atomic wave function (e.g., the one
given by the Wannier functions), which will lead to the modification
of both classical and quantum scattering. Moreover, we assumed the
uniform dependence of all fluctuations in space. Various spatial
dependences of the correlations as
$\langle\hat{n}_i\hat{n}_j\rangle(\bf{r}_i-\bf{r}_j)$ will manifest
themselves in the angular distribution of the scattered light. Here,
we presented the calculations of some expectation values, which
assumes the repeated measurement. To fully characterize the quantum
measurement process, taking into account the measurement back-action
should be done. A single-run measurement of scattered photons will
lead to the reduction of the atomic state as well, due to the
entanglement between light and matter \cite{arxiv}.

\newpage

Fig. 1. General setup. Atoms in a deep lattice (the trapping beams
are not shown) are illuminated by a probe wave $a_0$ at angle
$\theta_0$. Additional probing through a mirror $\eta$ is possible.
The scattered light $a_1$ is collected by a cavity at angle
$\theta_1$ and measured by a detector.

$$\\$$

Fig. 2. Transmission spectra of a traveling wave cavity. (a) Setup.
(b) Cavity photon numbers at different probe-cavity detunings.
Single Lorentzian for the Mott insulator state and many Lorentzians
for the superfluid state, $\kappa=0.1g^2/\Delta_{0a}$, $N=M=30$,
$K=15$.

$$\\$$

Fig. 3. Intensity angular distributions for two traveling waves, the
probe is transverse to the lattice ($\theta_0=0$), the lattice
period is $d=\lambda/2$. (a) Setup. (b) Intensity distribution of
the classical diffraction. (c) Noise quantity $R(\theta_1)$ for the
superfluid state. $N=M=K=30$.

$$\\$$

Fig. 4. Intensity angular distributions for scattering of a
traveling probe into a standing-wave cavity. The probe is at
$\theta_0=0.1\pi$. (a) Setup. (b) Intensity distribution of the
classical diffraction. (c) Noise quantity $R(\theta_1)$ for the
superfluid state. $N=M=K=30$, $d=\lambda/2$.

$$\\$$

Fig. 5. Intensity angular distributions for scattering of a
standing-wave probe into a standing-wave cavity. The probe is at
$\theta_0=0.1\pi$. (a) Setup. (b) Intensity distribution of the
classical diffraction. (c) Noise quantity $R(\theta_1)$ for the
superfluid state. $N=M=K=30$, $d=\lambda/2$.

\newpage

\begin{figure}
\scalebox{0.9}[0.9]{\includegraphics{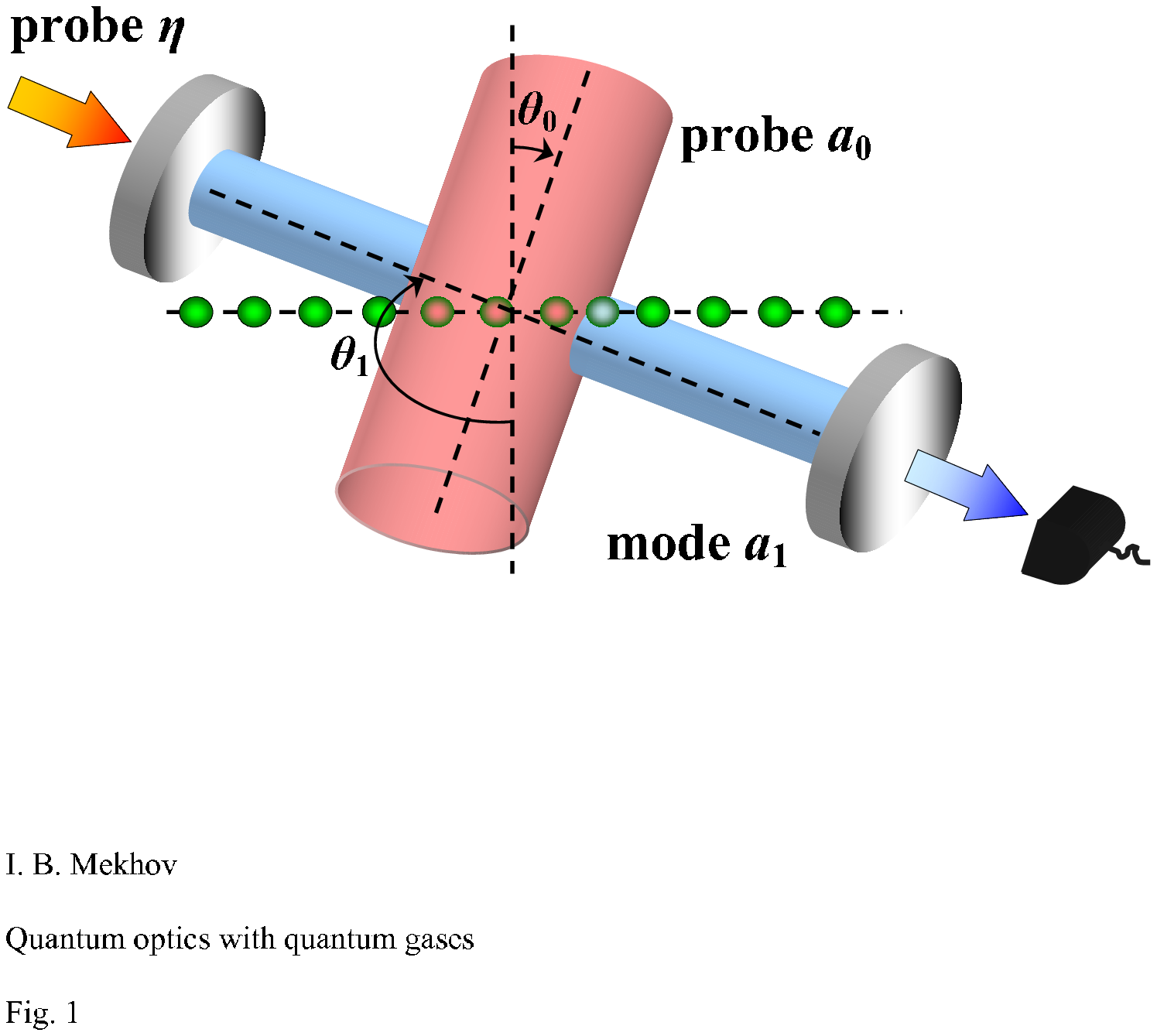}}
\end{figure}

\begin{figure}
\scalebox{0.9}[0.9]{\includegraphics{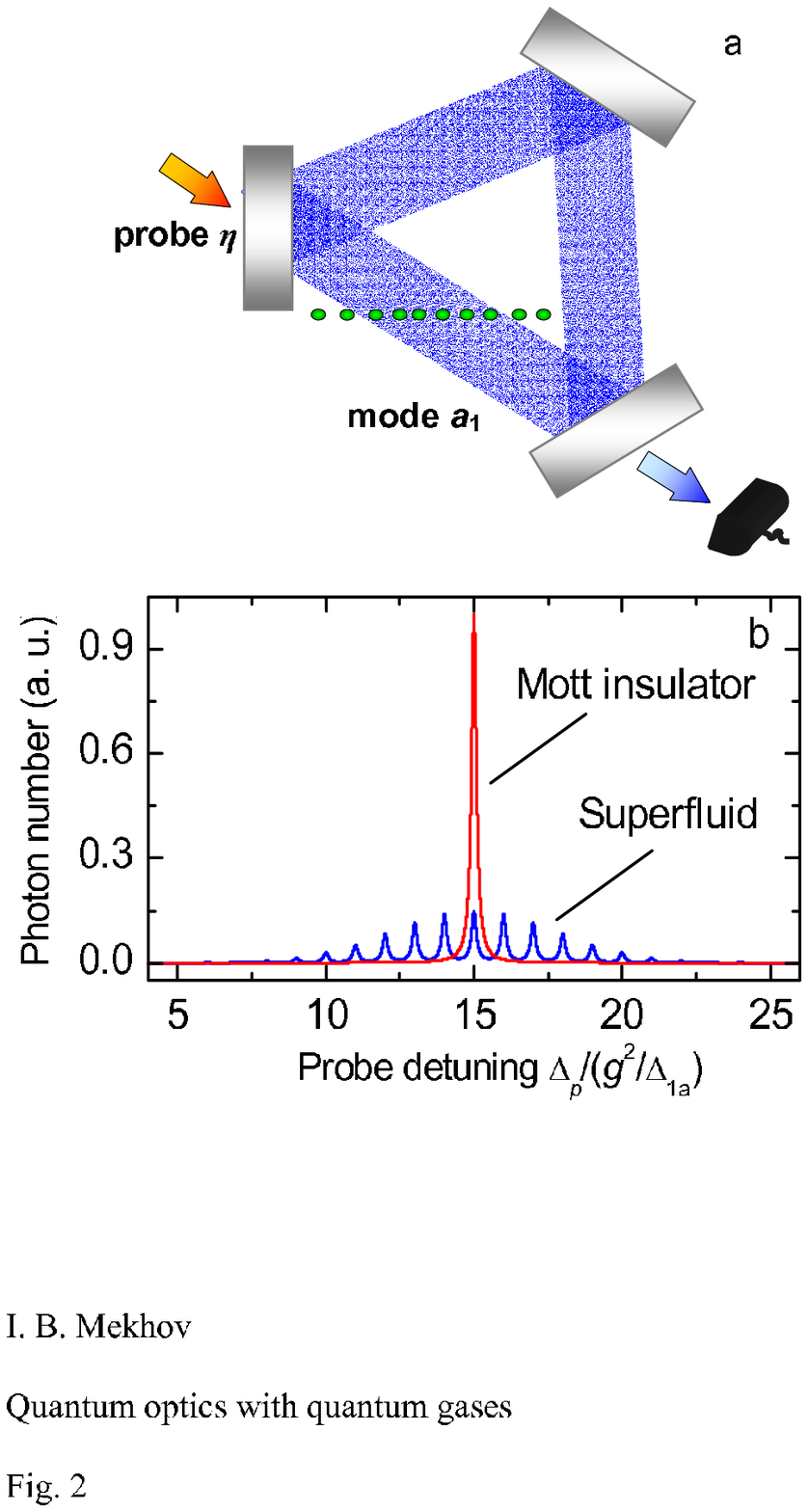}}
\end{figure}

\begin{figure}
\scalebox{0.9}[0.9]{\includegraphics{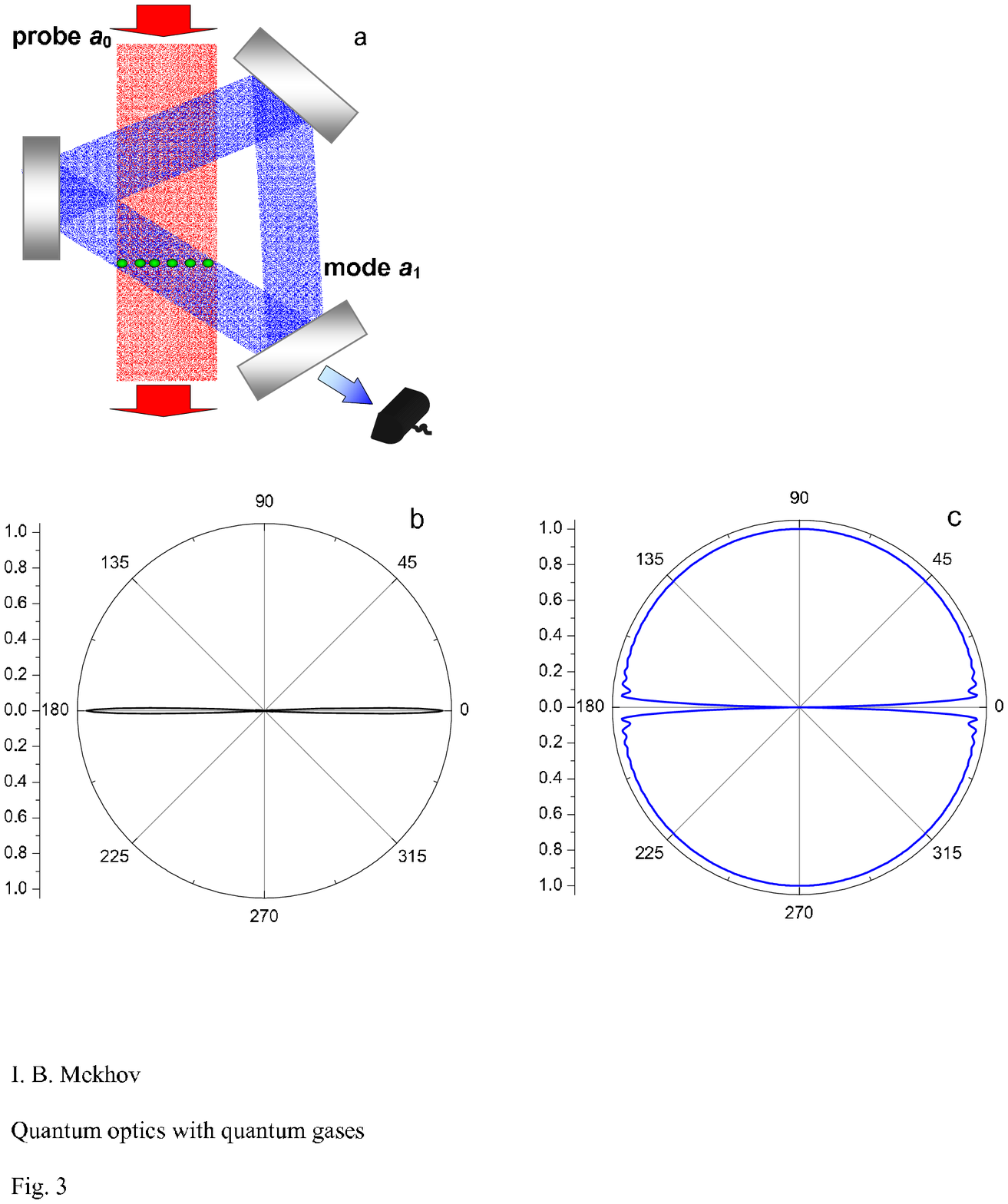}}
\end{figure}

\begin{figure}
\scalebox{0.9}[0.9]{\includegraphics{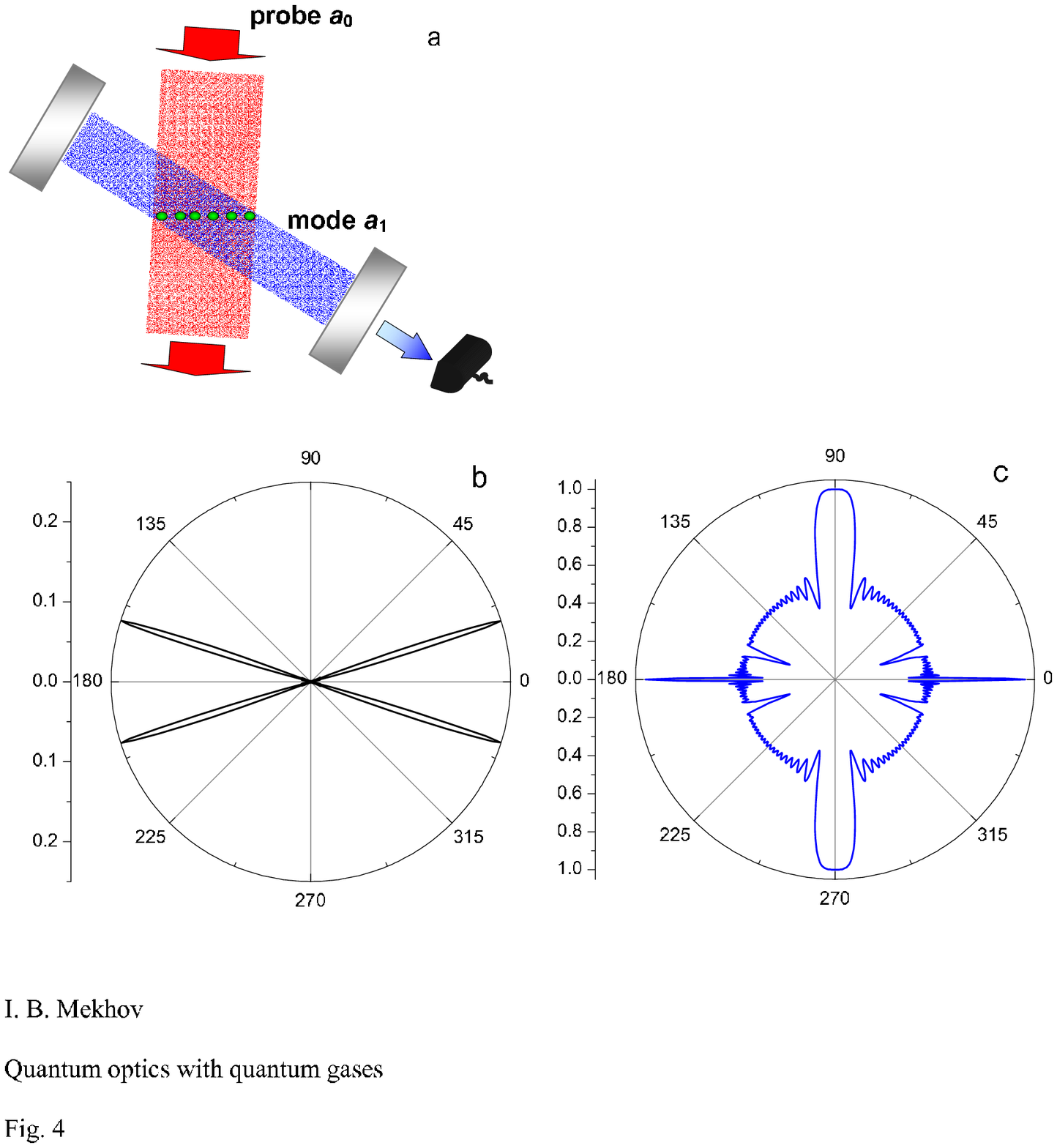}}
\end{figure}

\begin{figure}
\scalebox{0.9}[0.9]{\includegraphics{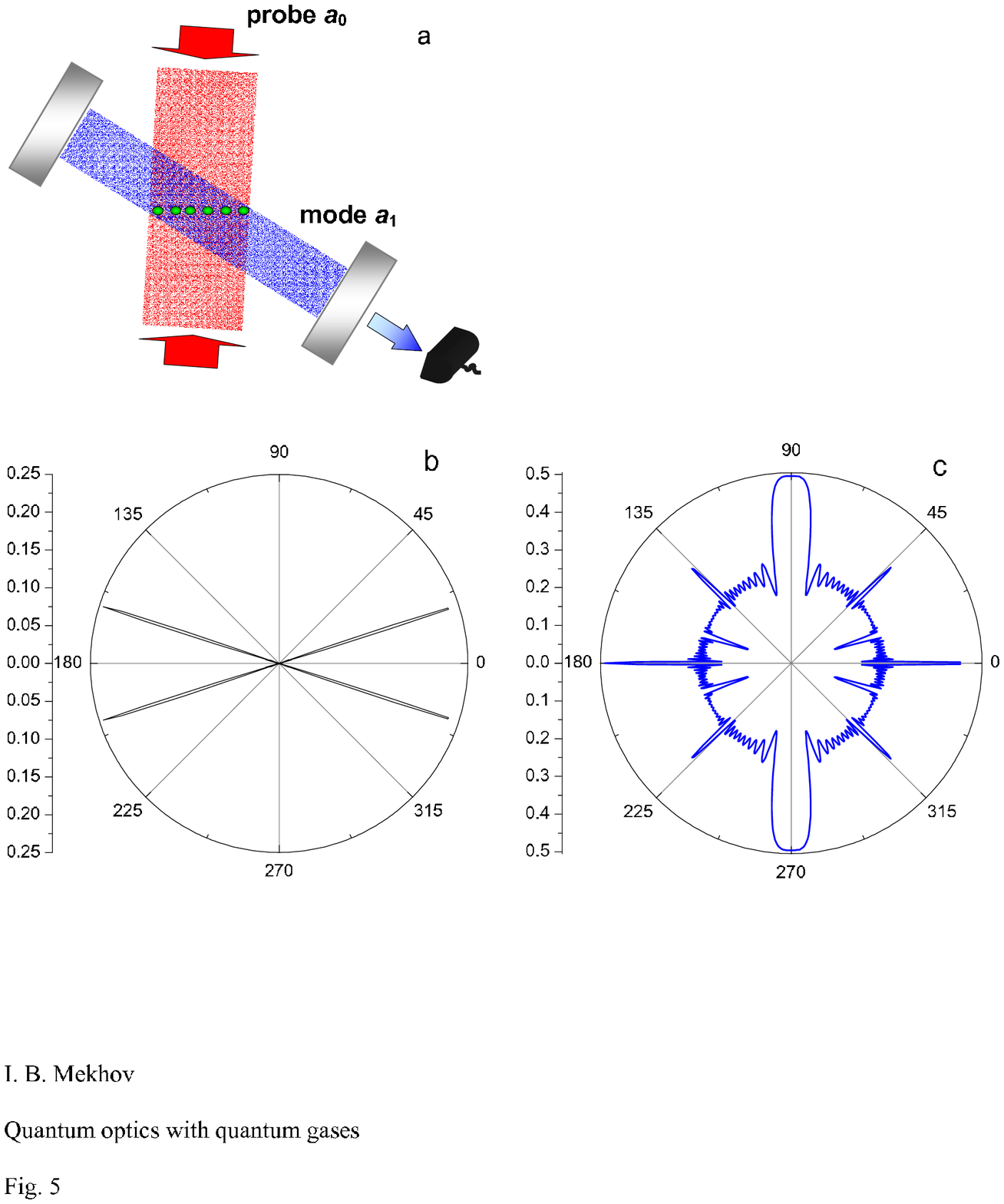}}
\end{figure}

\end{document}